\documentclass{article}
\usepackage{amsmath,epsfig}
\usepackage[preprint]{spconfa4}
\usepackage{times}  
\usepackage{helvet}  
\usepackage{courier}  
\usepackage{url}  
\usepackage{graphicx}  
\usepackage[font=scriptsize,labelfont=bf]{caption}
\usepackage[belowskip=-10pt,aboveskip=3pt]{caption}

\usepackage[font=scriptsize,labelfont=bf]{caption}
\usepackage[belowskip=-10pt,aboveskip=3pt]{caption}

\usepackage{multirow}
\usepackage{tabularx}
\usepackage{dsfont}
\usepackage{url}
\usepackage[tight]{subfigure}
\usepackage{color}
\usepackage{subfigure}
\usepackage{amsthm}
\usepackage[export]{adjustbox}
\usepackage{comment}

\usepackage[utf8]{inputenc} 
\usepackage[T1]{fontenc}    
\usepackage{url}            
\usepackage{booktabs}       
\usepackage{amsfonts}       
\usepackage{nicefrac}       
\usepackage{microtype}      

\def\argmin{\operatornamewithlimits{arg\,min}}
\def\argmax{\operatornamewithlimits{arg\,max}}

\usepackage{colortbl}
\definecolor{tabgray}{rgb}{0.85,0.85,0.85}
\definecolor{top1}{rgb}{1.0, 0.6, 0.6} 
\definecolor{top2}{rgb}{0.98, 0.91, 0.71}
\definecolor{top3}{rgb}{0.91, 1.0, 1.0}
\definecolor{top1-2}{rgb}{1.0, 0.66, 0.66} 
\definecolor{top1-3}{rgb}{1.0, 0.72, 0.72} 
\definecolor{top1-4}{rgb}{1.0, 0.78, 0.78} 
\definecolor{top1-5}{rgb}{1.0, 0.84, 0.84} 
\definecolor{top1-6}{rgb}{1.0, 0.90, 0.90} 
\definecolor{top1-7}{rgb}{1.0, 0.96, 0.96} 

\definecolor{dg}{rgb}{0,0.694,0.298}
\definecolor{purple}{rgb}{0.4,0.176,0.569}

\usepackage{pifont}
\usepackage{xspace}
\makeatletter
\DeclareRobustCommand\onedot{\futurelet\@let@token\@onedot}
\def\@onedot{\ifx\@let@token.\else.\null\fi\xspace}
\def\eg{\emph{e.g}\onedot} 
\def\ie{\emph{i.e}\onedot} 
 
\def\etc{\emph{etc}\onedot}

\makeatother

\definecolor{citecolor}{RGB}{65,105,225}
\usepackage[breaklinks=true,colorlinks,citecolor=citecolor,bookmarks=false]{hyperref}

\copyrightnotice{Copyright notice –
978-1-6654-3864-3/21/\$31.00~\copyright 2021 IEEE}

\let\OLDthebibliography\thebibliography
\renewcommand\thebibliography[1]{
  \OLDthebibliography{#1}
  \setlength{\parskip}{0pt}
  \setlength{\itemsep}{0pt plus 0.3ex}
}

\begin{document}\sloppy

\def\x{{\mathbf x}}
\def\L{{\cal L}}

\title{BIAS FIELD POSES A THREAT TO DNN-BASED X-RAY RECOGNITION}
%
\name{%
\begin{tabular}{@{}c@{}}
Binyu Tian$^{1}$ \quad 
Qing Guo$^{2^{\ast}}$ \quad 
Felix Juefei-Xu$^{3}$ \quad
Wen Le Chan$^{2}$ \quad 
Yupeng Cheng$^{2}$ \\
Xiaohong Li$^{1^{\ast}}$ \quad
Xiaofei Xie$^{2}$\quad
Shengchao Qin$^{4}$\thanks{$^{\ast}$Qing Guo and Xiaohong Li are the corresponding authors (tsingqguo@ieee.org and xiaohongli@tju.edu.cn).}
\end{tabular}}
%

\address{$^{1}$College of Intelligence and Computing, Tianjin University, China \\ \quad $^{2}$Nanyang Technological University, Singapore \\
$^{3}$Alibaba Group, USA \quad $^{4}$Teesside University, UK}

\maketitle

\begin{abstract}
Chest X-ray plays a key role in screening and diagnosis of many lung diseases including the COVID-19. 
Many works construct deep neural networks (DNNs) for chest X-ray images to realize automated and efficient diagnosis of lung diseases.
However, \textit{bias field} caused by the improper medical image acquisition process widely exists in the chest X-ray images while the robustness of DNNs to the bias field is rarely explored, posing a threat to the X-ray-based automated diagnosis system. 
In this paper, we study this problem based on the adversarial attack and propose a brand new attack, \ie, \textit{adversarial bias field attack} where the bias field instead of the additive noise works as the adversarial perturbations for fooling DNNs. 
This novel attack poses a key problem: how to locally tune the \textit{bias field} to realize high attack success rate while maintaining its spatial smoothness to guarantee high realisticity. These two goals contradict each other and thus has made the attack significantly challenging.
To overcome this challenge, we propose the \textit{adversarial-smooth bias field attack} that can locally tune the bias field with joint smooth \& adversarial constraints. As a result, the adversarial X-ray images can not only fool the DNNs effectively but also retain very high level of realisticity.
We validate our method on real chest X-ray datasets with powerful DNNs, \eg, ResNet50, DenseNet121, and MobileNet, and show different properties to the state-of-the-art attacks in both image realisticity and attack transferability.
Our method reveals the potential threat to the DNN-based X-ray automated diagnosis and can definitely benefit the development of bias-field-robust automated diagnosis system.
\end{abstract}
%
\section{Introduction}\label{sec:intro}






\begin{figure}[t]
	\centering
	\includegraphics[width=0.9\columnwidth]{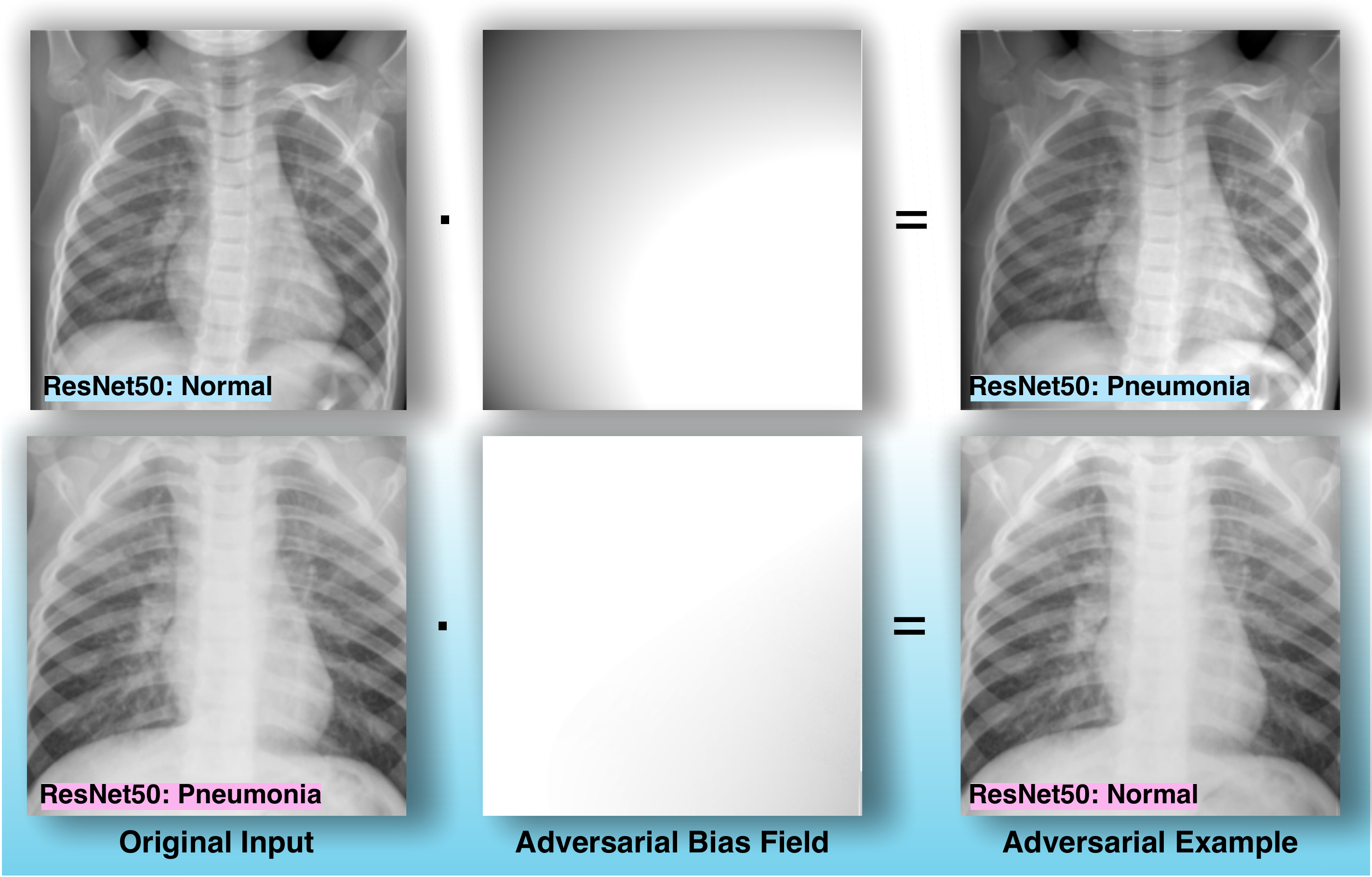}
	\caption{Two cases of our adversarial bias field examples. Our proposed adversarial-smooth bias field attack can adversarially but imperceptibly altered the bias field, misleading the advanced DNN models, \eg, ResNet50, to diagnose the normal X-ray image as the pneumonia one. More troubling, the DNN could be fooled to diagnose the pneumonia X-ray image as the normal one, having higher risk of delaying patients' treatment.}
	\label{fig:motivation}
\end{figure}
Medical image diagnosis and recognition is starting to be automated by DNNs with a clear advantage of being very efficient in diagnosing the disease outcomes. However, unlike human experts, such automated methods based on DNNs still have some caveats. For example, with the presence of image-level degradations during the image acquisition process, the recognition accuracy can be dramatically suppressed. Sometimes, such DNN-based medical image recognition system can even become entirely vulnerable when maliciously attacked by an adversary or an abuser that is financially incentivized. 


There are mainly two types of image perturbations or degradations in medical imagery: (1) image noise, and (2) image bias field. The image noise is primarily caused by the image sensor noise and the image bias field is caused by the spatial variations of radiation \cite{vovk2007review}, which is common among medical imaging, ranging from magnetic resonance imaging (MRI) \cite{ahmed2002modified}, computed tomography (CT) \cite{guo2017frequency}, to X-ray imaging, \etc. The bias field appears as the intensity inhomogeneity in the MRI, CT, or X-ray images. For consumer digital imaging, the bias field shows up as the illumination changes or vignetting effect. 


In this work, we want to reveal this vulnerability caused by image bias field. To the best of our knowledge, this is the very first attempt to adversarially perturb the bias field, in order to attack DNN-based X-ray recognition. Contrary to the additive noise-perturbation attack on DNN-based recognition systems, the attack on the bias field is multiplicative in nature \cite{zheng2010estimation}, which is fundamentally different from the noise attack. What is more important is that in order to make the bias field attack realistic and imperceptible, the successful attacks need to maintain the smoothness property of the bias field, which is genuinely more challenging because local smoothness usually contradicts with high attack success rates.


To overcome this challenge, we capitalize on this proprietary degradation surrounding X-ray imagery and initiate adversarial attacks based on imperceptible modification on the bias field itself. Specifically, we have proposed the adversarial-smooth bias field generator that can locally tune the bias field with joint smooth and adversarial constraints by tapping into the bias field generation process based on a multivariate polynomial model. As a result, the adversarially perturbed bias field applied to the X-ray image can not only fool the DNN-based recognition system effectively, but also retain high level of realisticity. We have validated our proposed method on several chest X-ray classification datasets with the state-of-the-art DNNs such as ResNet, DenseNet, and MobileNet, by showing superior performance in terms of both image realisticity and high attack success rates. A careful investigation into which bias field region contributes more significantly to the adversarial nature of the attack can lead to a better interpretation and understanding of the DNN-based recognition system and its vulnerability, which, we believe, is of utmost importance. The ultimate goal of this work is to reveal that the bias field does pose a potential threat to the DNN-based automated recognition system, and can definitely benefit the development of bias-field-robust automated diagnosis system in the future.



\section{Related work}\label{sec:related_work}
%

\textbf{X-Ray imagery recognition.}
%
%
%
X-ray radiography is widely used in the medical field for diagnosis or treatment of diseases. \cite{Wang2017CXRDataset} releases the ChestX-ray14 dataset and evaluates the performance of 4 classic convolutional neural network (CNN) on the multi-label image classification of diseases.  \cite{yao2017} proposes the use of a CNN backbone with a variant of DenseNet model. 
\cite{Guan2018} presents a three-branch attention guided CNN that combines local cues and global features. 
CNNs are also explored for COVID-19 detection \cite{COVIDwang2020,COVIDafshar}, motivated by the need of quick and convenient screening, since abnormalities can be found in some patients' chest X-Ray images. 

Despite considerations made to address data irregularities like class imbalance in dataset, the effect of medical image degradation is rarely addressed. For example, bias field could adversely affect quantitative image analysis \cite{Juntu2005BiasField}. Though many inhomogeneity correction strategy are proposed \cite{fan2003unified,thomas20053d}, the possible detrimental effect on disease identification, location or segmentation by bias field is rarely explored, possibly reducing DNN's robustness. To the best of authors' knowledge, this paper is very first work that looks at the effect of bias field from the view of adversarial attack.

\textbf{General adversarial attack.}
DNNs in image, speech or natural language processing application are susceptible to adversarial attacks \cite{goodfellow2014explaining,fredrikson2015model,papernot2017practical,carlini2017towards,guo2020abba,guo2020spark,wang2020amora}. 
Specifically, fast gradient sign method (FGSM) proposed by \cite{goodfellow2014explaining} involves only one back propagation step when calculating the cost function's gradient, allowing fast adversarial example generation. \cite{kurakin2016adversarial} proposes basic iteration method (BIM), an iterative version of FGSM.  \cite{carlini2017towards} proposes to use margin loss instead of entropy loss during attacks. \cite{papernot2017practical} proposes the exploitation of transferability of adversarial examples. 


\textbf{Adversarial attack on medical imagery.}
There are existing literature that look into adversarial attack against deep learning system for medical imagery. \cite{finlayson2018adversarial} shows that both black box and white box PGD attack and adversarial patch attack can affect the classifiers' performance on fundoscopy, chest X-ray and dermoscopy, respectively. By producing crafted mask, an adaptive segmentation mask attack (ASMA) is proposed to fool DNN model \cite{ozbulak2019impact}.


However, very few literature has leveraged on and conducted adversarial attack based on the inherent characteristics of the targeted medical imagery. For example, common noise degradation used for general adversarial attacks are rarely found in X-ray imagery. Hence in this work, we capitalize on the proprietary degradation surrounding X-ray imagery, bias field, and initiate adversarial attacks based on imperceptible modification on the bias field itself.
\begin{figure}[]
	\centering
	\includegraphics[width=0.9\columnwidth]{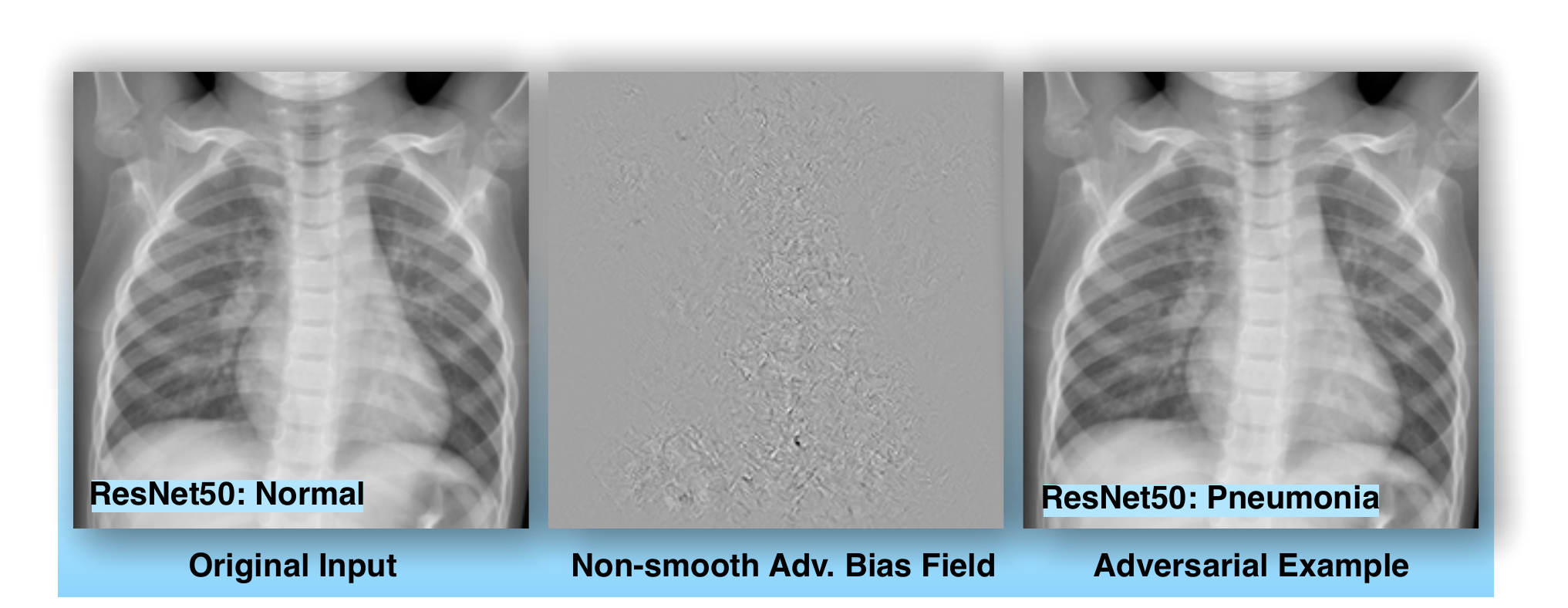}
	\caption{An example of using Eq.~\eqref{eq:biasfield_obj} to generate the non-smooth adversarial bias field.}
	\label{fig:nonsmbf}
\end{figure}




\section{Methodology}\label{sec:method}

\subsection{Adversarial Bias Field Attack and Challenges}
Given a X-ray image, \eg, $\mathbf{X}^\text{a}$, we can assume it is generated by adding a bias field $\mathbf{B}$ to a clean version, \ie, $\mathbf{X}$, with the widely used imaging model
%
\begin{align}\label{eq:biasfield}
\mathbf{X}^\mathrm{a}=\mathbf{X}\mathbf{B}.
\end{align}
%
Under the automate diagnosis task where a DNN is used to recognize the category (\ie, normal or abnormal) of $\mathbf{X}^\text{a}$, it is necessary to explore a totally new task, \ie, \textit{adversarial bias field attack} aiming to fool the DNN by calculating an adversarial bias field $\mathbf{B}$, with which we can study the influence of the bias field as well as the potential threat of utilizing it to fool the automate diagnosis.

A simple way is to take logarithm on Eq.~\eqref{eq:biasfield} and transform the multiplication to additive operation
%
\begin{align}\label{eq:biasfield_log}
\hat{\mathbf{X}}^\mathrm{a}=\hat{\mathbf{X}}+\hat{\mathbf{B}},
\end{align}
%
where we use the `$\hat{\cdot}$' to represent the logarithm of a variable. 
With Eq.~\eqref{eq:biasfield_log}, it seems that all existing additive-based adversarial attacks, \ie, FGSM, BIM, MIFGSM, DIM, and TIMIFGSM, could be used for the new attack. For example, we can calculate $\hat{\mathbf{B}}$ to realize attack by solving
%
\begin{align}\label{eq:biasfield_obj}
\argmax_{\hat{\mathbf{B}}} J(\hat{\mathbf{X}}+\hat{\mathbf{B}}, y), \text{~~subject to~~} \, \|\hat{\mathbf{B}}\|_\text{p}\leq\epsilon,
\end{align}
%
where $J(\cdot)$ is the loss function for classification, \eg, the cross-entropy loss, and $y$ denotes the ground truth label of $\mathbf{X}$.
Nevertheless,  we argue that such solution cannot generate the real `bias field' since the optimized $\hat{\mathbf{B}}$ violated the basic property of bias field, \ie, 
\textit{spatially smooth changes} resulting in intensity inhomogeneity. For example, as shown in Fig.~\ref{fig:nonsmbf}, when we optimize Eq.~\eqref{eq:biasfield_obj} to produce a bias field, we can attack the ResNet50 successfully while the bias field is noise-like and far from the appearance in the real world.

As a result, due to requirement of spatial smoothness of bias field, the \textit{adversarial bias field attack} posts a totally \textit{new challenge} to the field of adversarial attack:
how to generate the adversarial perturbation that can not only achieve high attack success rate but maintain its spatial smoothness for the realisticity of bias field. 
Actually, since the high attack success rate relies on the pixel-wise tunable perturbation and violates the smoothness requirement of bias field, the two constraints contradicts each other and make the \textit{adversarial bias field attack} significantly challenging.

\subsection{Adversarial-Smooth Bias Field Attack}
In this section, we propose the \textit{distortion-aware multivariate polynomial model} to represent the bias field whose inherit property guarantees the spatial smoothness of the bias field while the distortion helps achieve effective attack. Then, we define a new objective function for effective attack by combining the constraints of spatially smooth bias field, sparsity of the original image with the adversarial loss. Finally, we introduce the optimization method and attack algorithm. 
\begin{figure}[t]
	\centering
	\includegraphics[width=1.0\columnwidth]{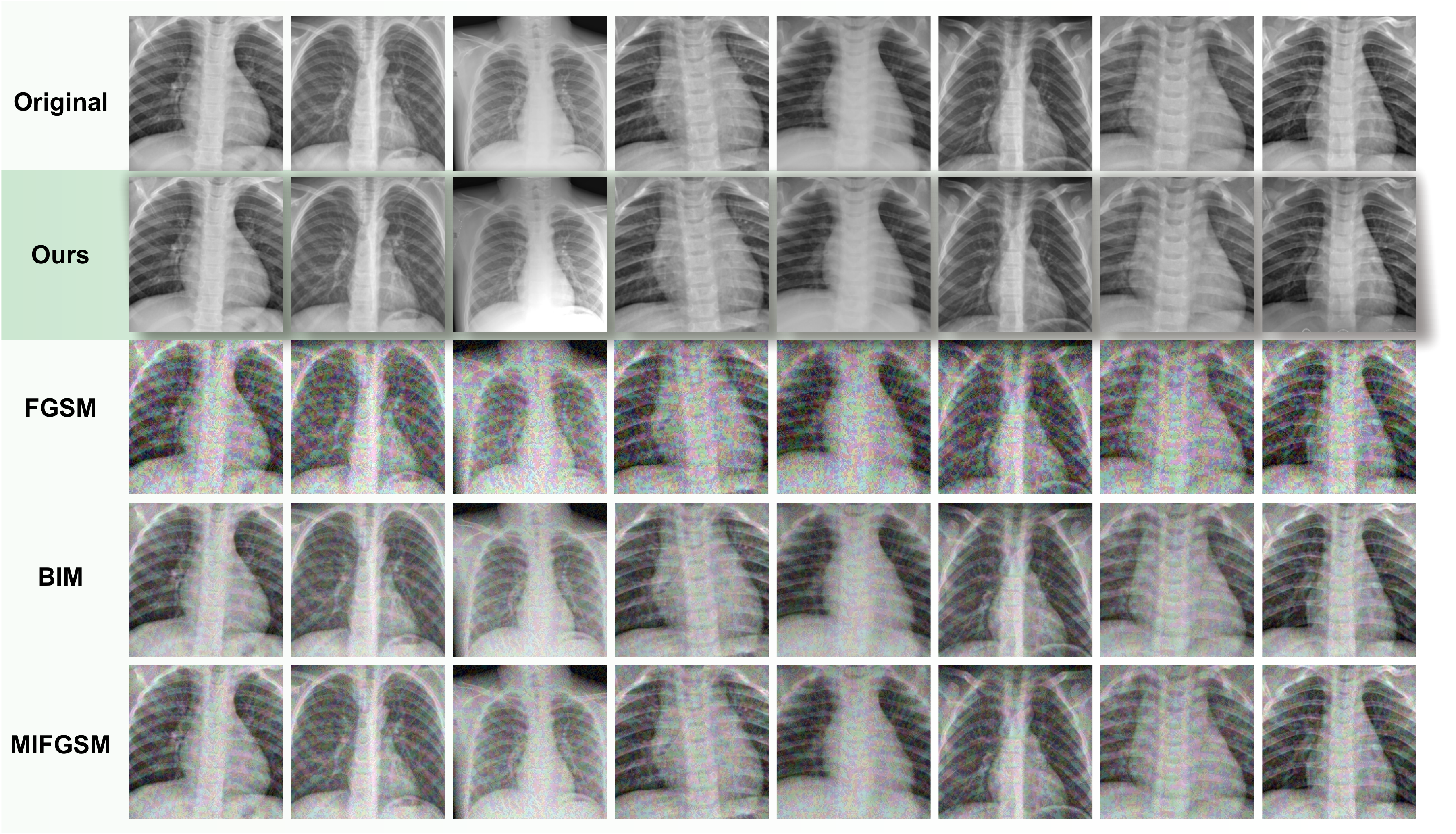}
	\caption{Examples of adversarial examples generated with different techniques.}
	\label{fig:visualization}
\end{figure}
\begin{figure}[t]
	\centering
	\includegraphics[width=0.8\columnwidth]{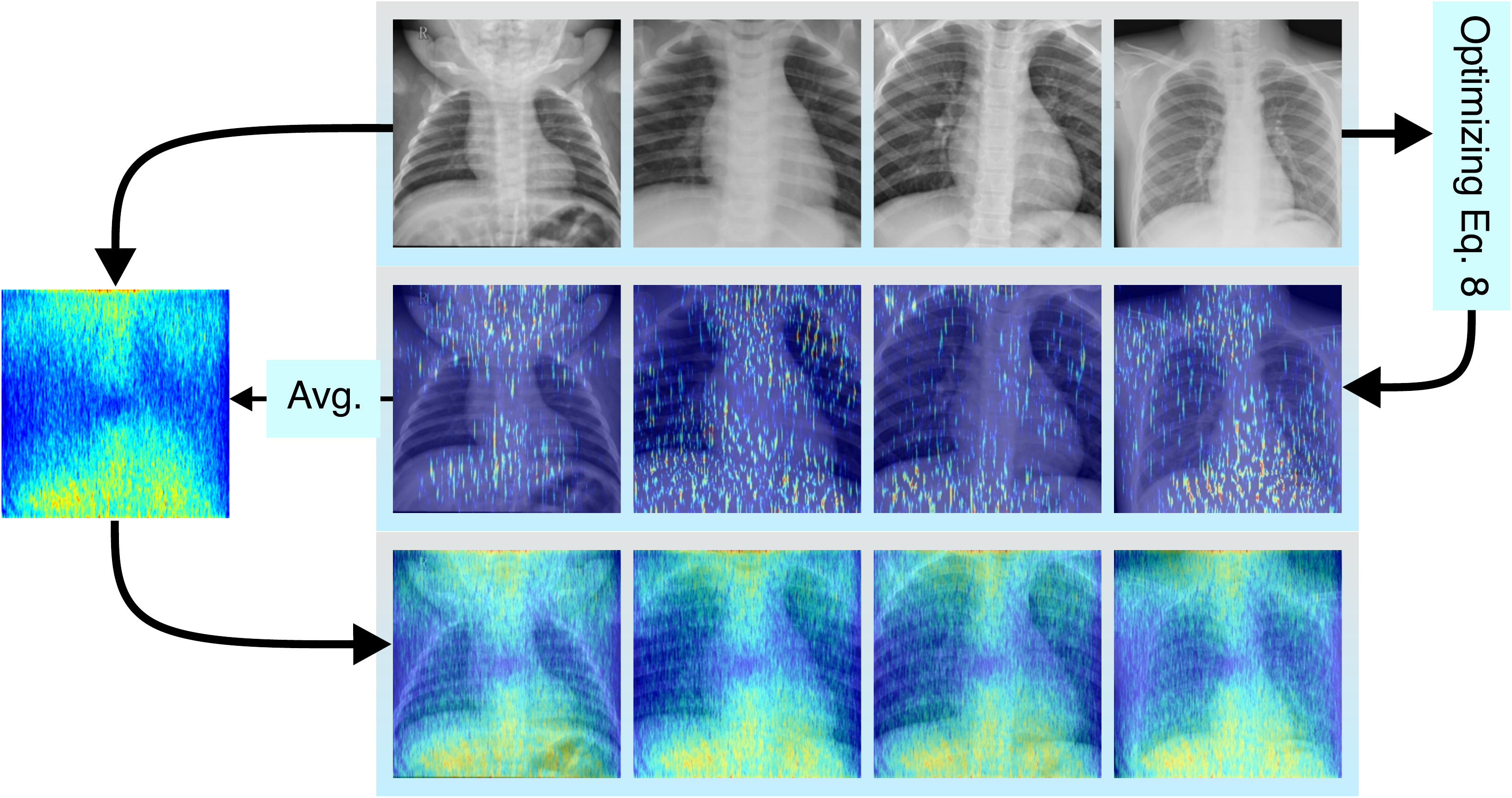}
	\caption{Pipeline and examples of exploring bias-field-sensitive regions. A subject model, \ie, ResNet50, is employed to generate adversarial bias field examples for 240 X-ray images and we then use Eq.~\eqref{eq:interp_map} to produce the interpretable map $\mathbf{M}$ for each image (\ie, the images at the second row where the maps are blended with the raw X-ray images for better understanding.). Finally, we can calculate an averaging map covering all interpretable maps and blend it with raw images (\ie, the images at the third row.)
	}
	\label{fig:analysis}
\end{figure}

{\bf Distortion-aware multivariate polynomial model.}
We model the bias filed $\hat{\mathbf{B}}$ as
%
\begin{align}\label{eq:biasfield_model}
\hat{\mathbf{B}}_i = \sum_{t=D_0}^{D}\sum_{l=D_0}^{D-t}{a_{t,l}\text{T}_\theta(x_i)^{t}\text{T}_\theta(y_i)^{l}}
\end{align}
%
where $\text{T}_\theta$ represents the distortion transformation and we use the thin plate spline (TPS) transformation with $\theta$ being the control points. We denote $i$ as the $i$-th pixel with its coordinates $(x_i,y_i)$ while $(\text{T}_\theta(x_i),\text{T}_\theta(y_i))$ means the pixel has been distorted by a TPS. In addition, $\{a_{t,l}\}$ and $D$ are the parameters and degree of the multivariate polynomial model, respectively, and the number of parameters are $|\{a_{t,l}\}|=\frac{(D-D_0+1)(D-D_0+2)}{2}$.
For convenient representations, we concatenate $\{a_{t,l}\}$ and obtain a vector $\mathbf{a}$.

{\bf Adversarial-smooth objective function.}
With Eq.~\eqref{eq:biasfield_model},  we can tune $\mathbf{a}$ and $\theta$ for adversarial attack and the multivariate polynomial model can help preserve the smoothness of bias field. 
%
Intuitively, on the one hand, the lower degree $D$ leads to less model parameters $|\{a_{t,l}\}|$ and smoother bias field. 
On the other hand, the distortion $(\text{T}_\theta(x_i),\text{T}_\theta(y_i))$ can be locally tuned with different $\theta$ and can help achieve effective attack. 
The key problem is how to calculate $\{a_{t,l}\}$ and $\theta$ to balance the spatial smoothness and adversarial attack. To this end, we define a new objective function to realize the attack.
%
\begin{align}\label{eq:biasfield_advobj}
\argmax_{\mathbf{a}, \theta}  J(\hat{\mathbf{X}}+\hat{\mathbf{B}}(\mathbf{a},\theta), y) - \lambda_a\|\mathbf{a}\|_1  - \lambda_\theta\|\theta-\theta_0\|_1,
\end{align}
%
where $\theta_0$ denotes parameters of the identify TPS transformation, \ie, $x_i=\text{T}_{\theta_0}(x_i)$. The first term is to tune the $\mathbf{a}$ and $\theta$ to fool a DNN. The second term encourages the sparse of $\{a_{t,l}\}$ and would let the bias field smooth. The final term is to let the TPS transformation not go far away from the identity version.
Two hyper-parameters, \ie, $\lambda_a$ and $\lambda_\theta$ control the balance between the smoothness and adversarial attack.

{\bf Optimization} \label{subsec:opt_alg}
Like the optimization methods used in general adversarial noise attack, we solve Eq.~\eqref{eq:biasfield_obj} and \eqref{eq:biasfield_advobj} via sign gradient descent where $\mathbf{a}$ and $\theta$ are updated via fixed rate
%
\begin{align}\label{eq:opt1}
\mathbf{a}_t=\mathbf{a}_{t-1}+\epsilon_a\text{sign}(\nabla{\mathbf{a}_{t-1}}), \\
\theta_t=\theta_{t-1}+\epsilon_\theta\text{sign}(\nabla{\theta_{t-1}}),
\end{align}
%
where $\nabla{\mathbf{a}_{t-1}}$ and  $\nabla{\theta_{t-1}}$ denote the gradient of $\mathbf{a}_{t-1}$ and $\theta_{t-1}$ with respect to the objective function in Eq.~\eqref{eq:biasfield_advobj}, respectively. For Eq.~\eqref{eq:biasfield_obj}, we use the same to update $\hat{\mathbf{B}}$ directly. We fix $\epsilon_a=\epsilon_\theta=0.06$ with the iteration number being 10.

\section{Experiments}\label{sec:exp}


\subsection{Setup and Dataset}
{\bf Dataset.}
We carry out our experiments on a chest-xray dataset about pneumonia, which contains 5863 X-ray images\footnote{\scriptsize{Please find more details about the dataset in \url{https://www.kaggle.com/paultimothymooney/chest-xray-pneumonia}.}}. These images were selected from retrospective cohorts. The dataset is divided into two categories, \ie, pneumonia and normal. 

\noindent{\bf Models.}
In order to show the effect of the attack on different models, we finetune three pre-trained models on the chest-xray dataset. The three models are ResNet50, MobileNet and Densenet121 (Dense121).The accuracy of ResNet50, MobileNet and Densenet121 is 88.62\%, 88.94\% and 87.82\%.

\noindent{\bf Metrics.}
We choose the attack success rate and image quality to evaluate the effectiveness of the bias field attack. The image quality measurement metric is BRISQUE \cite{mittal2012no}. BRISQUE is an unsupervised image quality assessment method. A high score for BRISQUE indicates poor image quality.

\noindent{\bf Baselines.}
We select five adversarial attack methods as our baselines, including basic iterative method (BIM) \cite{kurakin2016adversarial}, Carlini \& Wagner L2 method (C\&W$_\text{L2}$) \cite{carlini2017towards}, saliency map method (SaliencyMap) \cite{papernot2016limitations}, fast gradient sign method (FGSM) \cite{goodfellow2014explaining} and momentum iterative fast gradient sign method (MIFGSM) \cite{dong2018boosting}.
For the setup of hyperparameters of these baselines, we use the default setup of foolbox \cite{rauber2017foolbox}. We set max perturbation to be $\epsilon=0.1$ relative to [0,1] range in basic experiments. Besides, we set iterations as 10 for MIFGSM and BIM.
\begin{figure}[t]
	\centering
	\includegraphics[width=0.9\columnwidth]{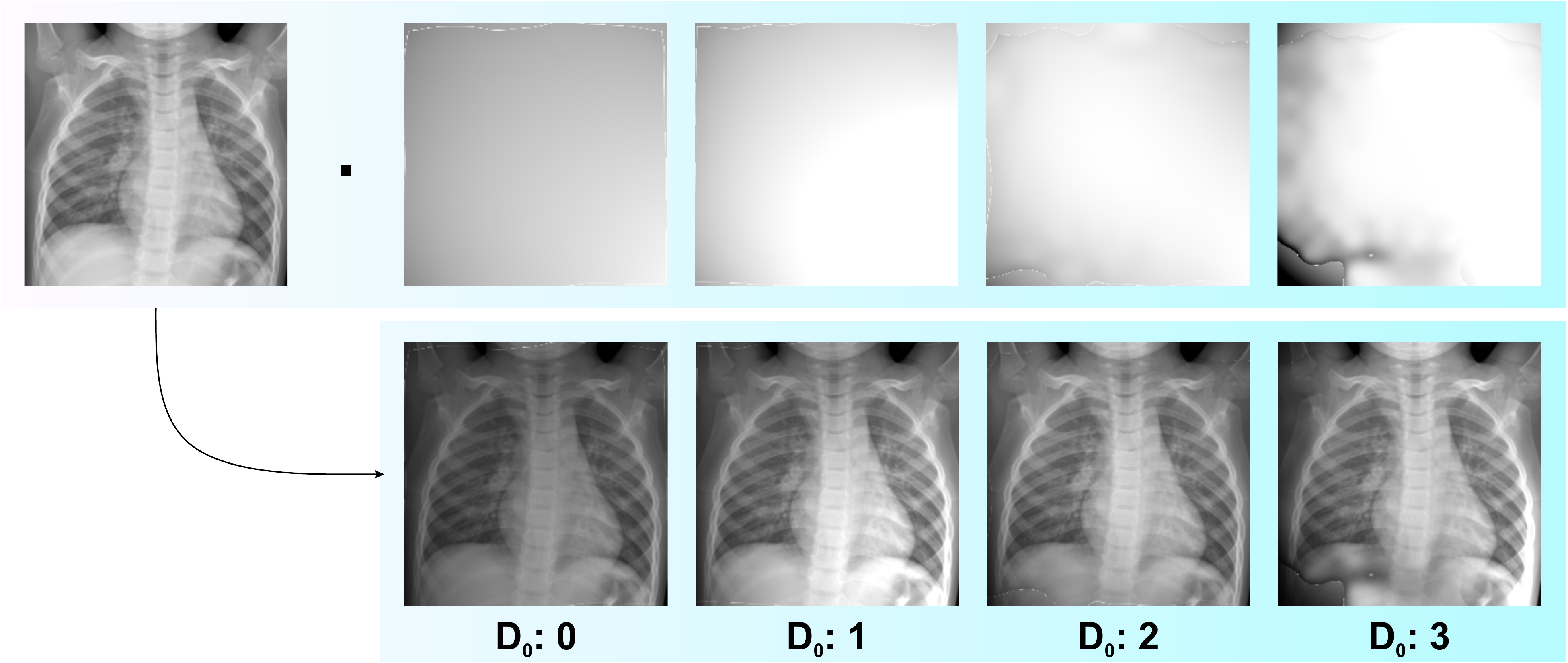}
	\caption{Effects of the multivariate polynomial model with different number of degrees, \ie, $D_0$ and $D$ in Eq.~\eqref{eq:biasfield_model}.}
	\label{fig:degree}
\end{figure}
\begin{table*}[t]
	\centering
	\small
	{
	\resizebox{1\linewidth}{!}{
		\begin{tabular}{l|cccc|cccc|cccc}
			
			\toprule
			
			\multicolumn{1}{c|}{Crafted from} & \multicolumn{4}{c|}{ResNet50} & \multicolumn{4}{c|}{Dense121} & \multicolumn{4}{c}{MobileNet} \\
			
			\cmidrule(r){1-1} \cmidrule(r){2-5} \cmidrule(r){6-9} \cmidrule(r){10-13}
			 Attacked model\&BRISQUE
			&  MobileNet &  Dense121  &  ResNet50 &BRISQUE
			
			&  ResNet50   &   MobileNet &  Dense121 &BRISQUE
			
			&  ResNet50   &   Dense121 &  MobileNet &BRISQUE\\
			
			\midrule
			BIM      &0.36	&0	&100 &30.0249  &0.54	&0.36 &100 &29.6599 &0	&0	&100   &29.9947   \\
			
			C\&W$_\text{L2}$
      &0.36	&0	&100	&30.1128 &1.08	&0.72	&100 &29.6455	&0	&0	&100	    &30.051  \\
      
			SaliencyMap
      &1.08	&0.18	&100
      &28.7108 &2.53	&1.26	&100	&28.4046	&0.72	&0.18	&100	&30.8351       \\

			FGSM &0	&0.18	&67.8
			&67.0028 &0.72	&0.72	&29.38 &28.5753	&0	&0	&30.09  &28.5404      \\

			MIFGSM      &0.36	&0	&100 &30.0578 &0.54	&0.36	&94.34 &29.6094 &0	&0	&100 &30.0134	       \\
			\hline
			
			AdvSBF (Ours)     & 7.57  & 14.05 & 38.69 &28.5703   &7.78   &  5.95 & 34.49 &28.9535 & 20.07 & 18.98 &  33.51     &29.5475   \\
			
			\bottomrule
			
		\end{tabular}
		}
	}
	\caption{Adversarial comparison results on chest-Xray dataset with five attack baselines and our method. It contains the success rates (\%) of transfer \& whitebox adversarial attack on three normally trained models: ResNet50, Dense121, and MobileNet. For each four columns, whitebox attack results are shown in the third one. The first two columns display the transfer attack results. And the last column shows the BRISQUE score.}
	\label{Tab-noisecompare}
\end{table*}
\begin{table*}[t]
	\centering
	\small
	{
	\resizebox{1\linewidth}{!}{
		\begin{tabular}{c|cccc|cccc|cccc}
			
			\toprule
			
		\multirow{2}{*}{$(grid size, grid size)$, $D_0$} & \multicolumn{4}{c|}{ResNet50} & \multicolumn{4}{c|}{Dense121} & \multicolumn{4}{c}{MobileNet} \\
			
		\cmidrule(r){2-5} \cmidrule(r){6-9} \cmidrule(r){10-13}
	
			&  MobileNet &  Dense121  &  ResNet50 &BRISQUE
			
			&  ResNet50   &   MobileNet &  Dense121 &BRISQUE
			
			&  ResNet50   &   Dense121 &  MobileNet &BRISQUE\\
			
			\midrule
			(4,4), 0
            &10.84	&15.33	&37.97	&32.4873	&14.65	&8.29	&31.39	&31.331
				&21.52	&20.44	&35.68	&34.9368

            \\
			(8,8), 0
            &9.91	&14.05	&37.79	&32.5778
            &13.2	&6.49	&31.57	&31.3609

            &21.7	&20.26	&35.68	&34.0957
            \\
            (12,12), 0
            &9.73	&14.23	&37.61	&32.097	&12.84	&6.49	&31.2	&31.9176	&21.7	&20.44	&35.86	&34.3194

            \\
            \hline
			(16,16), 0 &10.81	&14.42	&38.34	&32.3661	&13.56	&6.85	&31.02	&31.4455	&21.34	&20.26	&36.04	&34.0944
            \\
			(16,16), 1 &11.35	&13.5	&36.89	&31.3312	&14.65	&9.37	&32.12	&30.6853	&17	&19.34	&32.79	&31.7842
            \\
            (16,16), 2 &8.11	&7.85	&29.48	&29.0977	&12.84	&8.83	&26.09	&30.0223	&12.12	&11.86	&26.85	&29.5885\\
            (16,16), 3 &4.15	&2.19	&18.81	&28.606	&4.7	&4.68	&16.24	&29.0152	&4.7	&3.47	&15.32	&29.2909
\\
			
			\bottomrule
			
		\end{tabular}
		}
	}
	\caption{Adversarial comparison results on chest-Xray dataset with different setup of hyper-parameters in our method. It contains the success rates (\%) of transfer\& whitebox adversarial attacks. For each model, the first two columns display the blackbox attack results, the third one shows the attack results and the last column shows the BRISQUE score.}
	\label{Tab-noisecompare1}
\end{table*}

\subsection{Comparison with Baseline Methods}
For our method, we set the size of the control points, $D$ and $D_0$ as (16*16), 10, and 1, respectively.
Table~\ref{Tab-noisecompare} shows the quantitative results with our method and the baseline methods, which are conducted with different settings.  Specifically, we conduct two different attacks, i.e., the white-box attack and the transfer attack. The white-box attack aims to attack the target DNN directly while the transfer attack attacks the target DNN with the adversarial examples generated from other models. For example, for the transfer attack in Table~\ref{Tab-noisecompare}, the attack is performed on DNNs in the first row, and the generated adversarial examples are used to attack DNNs in the first two columns of the second row. 

As we can see, for the white-box attack (i.e., the third column for each model), we could find that the success rate of our method is lower than the existing baselines. For example, on ResNet50, our method achieves 38.69\% success rate while most of the baselines achieves 100\% success rate. The main reason is that the existing attacking techniques could add arbitrary noises on the image, which is not realistic. However, our method has a strict smooth limitation such that the generated adversarial examples look more realistic. 
As shown in Fig.~\ref{fig:visualization}, we show some examples generated by different attacks. The first row shows the original images while the following rows list the corresponding adversarial examples. It is clear that our method could generate high-quality adversarial examples that are smooth and realistic. In most cases, the change between original image and the generated image is imperceptible. However, we could find obvious noises in the adversarial examples generated by the baseline methods. Such noises are difficult to appear in X-rays in the real world.

For the transfer attack (\ie, the first two columns), we found that our method achieves much higher success rate than others. For example, the attack on ResNet50 achieves 7.57\% and 14.05\% transfer success rate on MobileNet and DenseNet121, respectively. However, the the best results of the baseline are only 1.08\% and 0.18\%. It is because that existing techniques calculate the ad-hoc noise, which may be only effective on the target DNN but not on other models. However, our attack considers the smoothness such that the generated adversarial examples are more realistic. Such adversarial examples are more robust and could reveal the common weakness of different DNNs (\ie, higher success rate of the transfer attack). The results indicate that our method could generate high-quality adversarial examples.
We also compare the image quality with the BRISQUE score (\ie, the forth column). The results show that our method could achieve competitive results with the-state-of-the-arts.

In summary, our method aims to generate high-quality and realistic adversarial examples. To generate such adversarial examples, the attack success rate is naturally lower than the noise-based adversarial attack techniques.


\subsection{Understanding Effects of Bias Field}
In this subsection, we aim to explore how the bias field affect the DNN-based X-ray recognition. 
\cite{FongICCV2017} proposes a method for understanding DNNs with the adversarial noise attack and generates an interpretable map indicating the classification-sensitive regions of a DNN. Inspired this idea, we can study which regions in the chest X-ray images are sensitive to the bias filed and affect the X-ray recognition. Specifically, given an adversarial bias field example $\mathbf{X}^{\text{a}}$ generated by our method and the original image $\mathbf{X}$, we can calculate an interpretable map $\mathbf{M}$ for a DNN $\text{DNN}(\cdot)$ by optimizing
%
\begin{align}\label{eq:interp_map}
\argmin_{\mathbf{M}}~~\text{DNN}_y(\mathbf{M} \odot \mathbf{X}^\text{a} + (1-\mathbf{M}) \odot \mathbf{X}) \\~~
+ \lambda_{1} \| \mathbf{M}\|_{1} + \lambda_{2}\mathrm{TV} (\mathbf{M}) \nonumber
\end{align}
%
where $\text{DNN}_y(\cdot)$ denotes the score at label $y$ that is the ground truth label of $\mathbf{X}$ and $\mathrm{TV}(\cdot)$ is the total-variation norm. Intuitively, optimizing Eq.~(\ref{eq:interp_map}) is to find the region that causes misclassification. We optimize Eq.~(\ref{eq:interp_map}) via gradient decent in 150 iterations and fix $\lambda_{1}=0.05$ and $\lambda_{2}=0.2$. 

With Eq.~\eqref{eq:interp_map}, given a pre-trained model, \ie, $\text{DNN}(\cdot)$, and a dataset $\mathcal{X}$ containing the successfully attacked X-ray images, we calculate a $\mathbf{M}$ for each X-ray image and then average all interpretable maps to show the statistical regions that are sensitive to the bias field. For example, we adopt ResNet50 as the subject model and construct $\mathcal{X}$ with 240 attacked X-ray images that can fool ResNet50 successfully. Then, we calculate the interpretable maps for all images in $\mathcal{X}$ (\eg, the second row in Fig.~\ref{fig:analysis}) and average them, achieving a 
statistical mean map (\eg, the left image shown in Fig.~\ref{fig:analysis}). With the visualization results, we observe that: \ding{182} Our method helps identify the bias-field-sensitive regions in each attacked image and we observe that these regions are related to the organ positions. \textit{This demonstrates that the effects of the bias field to the DNN stems from intensity variation around organs.} \ding{183} According to the statistical mean map, we see that \textit{the bias-field sensitive regions mainly locate at the top and bottom positions across all attacked images}, suggesting that future designed DNN should consider the spatial variations within in X-ray images. We observe similar results on other DNNs (Please find more results in the supplementary material), hinting that these are common phenomenons in the DNN-based X-ray recognition and demonstrating the potential applications of this work.

\subsection{Effects of Hyper-parameters}

We also evaluate the effects of hyper-parameters in our attack, \ie, $\theta$ and $D$ in Equation~\ref{eq:biasfield_model}. 
Specifically, we change $\theta$ for TPS transformation by changing the number of control points. $(grid size \times grid size)$ is denoted to represent the control points in the TPS transformation. Then we select different $grid size$ to conduct the attack. For the parameter $D$, we set the fixed $D$ as 10 and change the value of $D_0$, \ie, observe part of the sample display of the bias field by ignoring the lowest $D_0$ degree in the multivariate polynomial model.

Table~\ref{Tab-noisecompare1} shows the results with different configurations. In the second row, we fix the $D_0$ as 0 and change value of $gridsize$ as 4, 8, 12 and 16, respectively. As we can see, there seems to be no clear difference in the attack success rate when the parameter $gridsize$ varies. We conjecture that the attack could easily reach the upper bound in terms of the success rate with different $gridsize$. 

Then we fix the $gridsize$ as 16 and change the parameter $D_0$ as 0, 1, 2 and 3 (in the third row).
As we can see, as $D_0$ increases (\ie, more lower degree are ignored), the success rate of our method decreases and the BRISQUE score decreases. 
It is reasonable as ignoring more low degree in Equation~\ref{eq:biasfield_model} may reduce the space of the manipulation,
resulting in higher image quality and lower attack success rate.
The visualization results are shown in Fig.~\ref{fig:degree}. When more lower degree is ignored (\ie, larger $D_0$), the bias field samples tend to be less smooth.


\section{Conclusion}\label{sec:concl}

Deep learning has been used in chest X-ray image recognition for the diagnosis of lung diseases (\eg, COVID-19). It is especially important to ensure the robustness of the DNN in this scenario. To tackle this problem, this paper proposed a new adversarial bias field attack, which aims to generate more realistic adversarial examples by adding more smooth perturbations instead of noises. We demonstrated the effectiveness of our attack on the widely used DNNs. The results show that our method can generate high quality adversarial examples, which achieve high success rate of the transfer attack. The generated realistic images can reveal issues of the DNN, which calls for the  attention of robustness enhancement of the deep learning-based healthcare system.

In the future, we will extend the proposed attack against other tasks, \eg, visual object tracking \cite{guo2020selective,guo2017structure,zhou2017selective} and DeepFake evasion \cite{huang2020fakepolisher,juefei2021countering}, and also in tandem with other natural modalities such as \cite{arxiv20_cosal,arxiv20_retinopathy,arxiv20_advrain}. In addition, we can regard the adversarial bias field as a new kind of mutation for DNN testing \cite{issta19_deephunter,issre18_mutation,du2019deepstellar,ase18_gauge}.  

\textbf{Acknowledgement.}
This work has partially been sponsored by the National Science Foundation of China (No. 61872262) and the Natural Science Foundation of Tianjin (No. KJZ40420200017).



\scriptsize
\bibliographystyle{IEEEbib}
\bibliography{ref}

\begin{thebibliography}{10}

\bibitem{vovk2007review}
Uro Vovk, Franjo Pernus, and Botjan Likar,
\newblock ``A review of methods for correction of intensity inhomogeneity in
  mri,''
\newblock {\em IEEE transactions on medical imaging}, vol. 26, no. 3, pp.
  405--421, 2007.

\bibitem{ahmed2002modified}
Mohamed~N Ahmed, Sameh~M Yamany, Nevin Mohamed, Aly~A Farag, and Thomas
  Moriarty,
\newblock ``A modified fuzzy c-means algorithm for bias field estimation and
  segmentation of mri data,''
\newblock {\em IEEE transactions on medical imaging}, vol. 21, no. 3, pp.
  193--199, 2002.

\bibitem{guo2017frequency}
Qing Guo, Shuifa Sun, Fangmin Dong, Wei Feng, Bruce~Zhi Gao, and Siyu Ma,
\newblock ``Frequency-tuned acm for biomedical image segmentation,''
\newblock in {\em 2017 IEEE International Conference on Acoustics, Speech and
  Signal Processing (ICASSP)}. IEEE, 2017, pp. 821--825.

\bibitem{zheng2010estimation}
Yuanjie Zheng and James~C Gee,
\newblock ``Estimation of image bias field with sparsity constraints,''
\newblock in {\em 2010 IEEE Computer Society Conference on Computer Vision and
  Pattern Recognition}. IEEE, 2010, pp. 255--262.

\bibitem{Wang2017CXRDataset}
Xiaosong Wang, Yifan Peng, Le~Lu, Zhiyong Lu, Mohammadhadi Bagheri, and
  Ronald~M. Summers,
\newblock ``Chestx-ray8: Hospital-scale chest x-ray database and benchmarks on
  weakly-supervised classification and localization of common thorax
  diseases,''
\newblock {\em CoRR}, vol. abs/1705.02315, 2017.

\bibitem{yao2017}
Li~Yao, Eric Poblenz, Dmitry Dagunts, Ben Covington, Devon Bernard, and Kevin
  Lyman,
\newblock ``Learning to diagnose from scratch by exploiting dependencies among
  labels,''
\newblock {\em CoRR}, vol. abs/1710.10501, 2017.

\bibitem{Guan2018}
Qingji Guan, Yaping Huang, Zhun Zhong, Zhedong Zheng, Liang Zheng, and Yi~Yang,
\newblock ``Diagnose like a radiologist: Attention guided convolutional neural
  network for thorax disease classification,''
\newblock {\em CoRR}, vol. abs/1801.09927, 2018.

\bibitem{COVIDwang2020}
Linda Wang and Alexander Wong,
\newblock ``Covid-net: A tailored deep convolutional neural network design for
  detection of covid-19 cases from chest x-ray images,''
\newblock {\em arXiv preprint arXiv:2003.09871}, 2020.

\bibitem{COVIDafshar}
Parnian Afshar, Shahin Heidarian, Farnoosh Naderkhani, Anastasia Oikonomou,
  Konstantinos~N Plataniotis, and Arash Mohammadi,
\newblock ``Covid-caps: A capsule network-based framework for identification of
  covid-19 cases from x-ray images,''
\newblock {\em arXiv preprint arXiv:2004.02696}, 2020.

\bibitem{Juntu2005BiasField}
Jaber Juntu, Jan Sijbers, Dirk Van~Dyck, and Jan Gielen,
\newblock ``Bias field correction for mri images,''
\newblock in {\em Computer Recognition Systems}, pp. 543--551. Springer, 2005.

\bibitem{fan2003unified}
Ayres Fan, William~M Wells, John~W Fisher, M{\"u}jdat Cetin, Steven Haker,
  Robert Mulkern, Clare Tempany, and Alan~S Willsky,
\newblock ``A unified variational approach to denoising and bias correction in
  mr,''
\newblock in {\em Biennial international conference on information processing
  in medical imaging}. Springer, 2003, pp. 148--159.

\bibitem{thomas20053d}
David~L Thomas, Enrico De~Vita, Ralf Deichmann, Robert Turner, and Roger~J
  Ordidge,
\newblock ``3d mdeft imaging of the human brain at 4.7 t with reduced
  sensitivity to radiofrequency inhomogeneity,''
\newblock {\em Magnetic Resonance in Medicine: An Official Journal of the
  International Society for Magnetic Resonance in Medicine}, vol. 53, no. 6,
  pp. 1452--1458, 2005.

\bibitem{goodfellow2014explaining}
Ian~J Goodfellow, Jonathon Shlens, and Christian Szegedy,
\newblock ``Explaining and harnessing adversarial examples,''
\newblock {\em arXiv preprint arXiv:1412.6572}, 2014.

\bibitem{fredrikson2015model}
Matt Fredrikson, Somesh Jha, and Thomas Ristenpart,
\newblock ``Model inversion attacks that exploit confidence information and
  basic countermeasures,''
\newblock in {\em Proceedings of the 22nd ACM SIGSAC Conference on Computer and
  Communications Security}, 2015, pp. 1322--1333.

\bibitem{papernot2017practical}
Nicolas Papernot, Patrick McDaniel, Ian Goodfellow, Somesh Jha, Z~Berkay Celik,
  and Ananthram Swami,
\newblock ``Practical black-box attacks against machine learning,''
\newblock in {\em Proceedings of the 2017 ACM on Asia conference on computer
  and communications security}, 2017, pp. 506--519.

\bibitem{carlini2017towards}
Nicholas Carlini and David Wagner,
\newblock ``Towards evaluating the robustness of neural networks,''
\newblock in {\em 2017 ieee symposium on security and privacy (sp)}. IEEE,
  2017, pp. 39--57.

\bibitem{guo2020abba}
Qing Guo, Felix Juefei-Xu, Xiaofei Xie, Lei Ma, Jian Wang, Wei Feng, and Yang
  Liu,
\newblock ``Abba: Saliency-regularized motion-based adversarial blur attack,''
\newblock {\em arXiv preprint arXiv:2002.03500}, 2020.

\bibitem{guo2020spark}
Qing Guo, Xiaofei Xie, Felix Juefei-Xu, Lei Ma, Zhongguo Li, Wanli Xue, Wei
  Feng, and Yang Liu,
\newblock ``Spark: Spatial-aware online incremental attack against visual
  tracking,''
\newblock in {\em Proceedings of the European Conference on Computer Vision
  (ECCV)}, 2020.

\bibitem{wang2020amora}
Run Wang, Felix Juefei-Xu, Xiaofei Xie, Lei Ma, Yihao Huang, and Yang Liu,
\newblock ``Amora: Black-box adversarial morphing attack,''
\newblock in {\em ACM Multimedia Conference (ACMMM)}, 2020.

\bibitem{kurakin2016adversarial}
Alexey Kurakin, Ian Goodfellow, and Samy Bengio,
\newblock ``Adversarial machine learning at scale,''
\newblock {\em arXiv preprint arXiv:1611.01236}, 2016.

\bibitem{finlayson2018adversarial}
Samuel~G Finlayson, Hyung~Won Chung, Isaac~S Kohane, and Andrew~L Beam,
\newblock ``Adversarial attacks against medical deep learning systems,''
\newblock {\em arXiv preprint arXiv:1804.05296}, 2018.

\bibitem{ozbulak2019impact}
Utku Ozbulak, Arnout Van~Messem, and Wesley De~Neve,
\newblock ``Impact of adversarial examples on deep learning models for
  biomedical image segmentation,''
\newblock in {\em International Conference on Medical Image Computing and
  Computer-Assisted Intervention}. Springer, 2019, pp. 300--308.

\bibitem{mittal2012no}
Anish Mittal, Anush~Krishna Moorthy, and Alan~Conrad Bovik,
\newblock ``No-reference image quality assessment in the spatial domain,''
\newblock {\em IEEE Transactions on image processing}, vol. 21, no. 12, pp.
  4695--4708, 2012.

\bibitem{papernot2016limitations}
Nicolas Papernot, Patrick McDaniel, Somesh Jha, Matt Fredrikson, Z~Berkay
  Celik, and Ananthram Swami,
\newblock ``The limitations of deep learning in adversarial settings,''
\newblock in {\em 2016 IEEE European symposium on security and privacy
  (EuroS\&P)}. IEEE, 2016, pp. 372--387.

\bibitem{dong2018boosting}
Yinpeng Dong, Fangzhou Liao, Tianyu Pang, Hang Su, Jun Zhu, Xiaolin Hu, and
  Jianguo Li,
\newblock ``Boosting adversarial attacks with momentum,''
\newblock in {\em Proceedings of the IEEE conference on computer vision and
  pattern recognition}, 2018, pp. 9185--9193.

\bibitem{rauber2017foolbox}
Jonas Rauber, Wieland Brendel, and Matthias Bethge,
\newblock ``Foolbox: A python toolbox to benchmark the robustness of machine
  learning models,''
\newblock {\em arXiv preprint arXiv:1707.04131}, 2017.

\bibitem{FongICCV2017}
R.~C. {Fong} and A.~{Vedaldi},
\newblock ``Interpretable explanations of black boxes by meaningful
  perturbation,''
\newblock in {\em ICCV}, 2017, pp. 3449--3457.

\bibitem{guo2020selective}
Qing Guo, Ruize Han, Wei Feng, Zhihao Chen, and Liang Wan,
\newblock ``Selective spatial regularization by reinforcement learned decision
  making for object tracking,''
\newblock {\em IEEE Transactions on Image Processing}, vol. 29, pp. 2999--3013,
  2020.

\bibitem{guo2017structure}
Qing Guo, Wei Feng, Ce~Zhou, Chi-Man Pun, and Bin Wu,
\newblock ``Structure-regularized compressive tracking with online data-driven
  sampling,''
\newblock {\em IEEE Transactions on Image Processing}, vol. 26, no. 12, pp.
  5692--5705, 2017.

\bibitem{zhou2017selective}
Ce~Zhou, Qing Guo, Liang Wan, and Wei Feng,
\newblock ``Selective object and context tracking,''
\newblock in {\em 2017 IEEE International Conference on Acoustics, Speech and
  Signal Processing (ICASSP)}. IEEE, 2017, pp. 1947--1951.

\bibitem{huang2020fakepolisher}
Yihao Huang, Felix Juefei-Xu, Run Wang, Qing Guo, Lei Ma, Xiaofei Xie, Jianwen
  Li, Weikai Miao, Yang Liu, and Geguang Pu,
\newblock ``Fakepolisher: Making deepfakes more detection-evasive by shallow
  reconstruction,''
\newblock in {\em Proceedings of the 28th ACM International Conference on
  Multimedia}, 2020, pp. 1217--1226.

\bibitem{juefei2021countering}
Felix Juefei-Xu, Run Wang, Yihao Huang, Qing Guo, Lei Ma, and Yang Liu,
\newblock ``Countering malicious deepfakes: Survey, battleground, and
  horizon,''
\newblock {\em arXiv:2103.00218}, 2021.

\bibitem{arxiv20_cosal}
Ruijun Gao, , Qing Guo, Felix Juefei-Xu, Hongkai Yu, Xuhong Ren, Wei Feng, and
  Song Wang,
\newblock ``{Making Images Undiscoverable from Co-Saliency Detection},''
\newblock {\em arXiv preprint arXiv:2009.09258}, 2020.

\bibitem{arxiv20_retinopathy}
Yupeng Cheng, Felix Juefei-Xu, Qing Guo, Huazhu Fu, Xiaofei Xie, Shang-Wei Lin,
  Weisi Lin, and Yang Liu,
\newblock ``{Adversarial Exposure Attack on Diabetic Retinopathy Imagery},''
\newblock {\em arXiv preprint arXiv:2009.09231}, 2020.

\bibitem{arxiv20_advrain}
Liming Zhai, Felix Juefei-Xu, Qing Guo, Xiaofei Xie, Lei Ma, Wei Feng,
  Shengchao Qin, and Yang Liu,
\newblock ``{It's Raining Cats or Dogs? Adversarial Rain Attack on DNN
  Perception},''
\newblock {\em arXiv preprint arXiv:2009.09205}, 2020.

\bibitem{issta19_deephunter}
Xiaofei Xie, Lei Ma, Felix Juefei-Xu, Minhui Xue, Hongxu Chen, Yang Liu,
  Jianjun Zhao, Bo~Li, Jianxiong Yin, and Simon See,
\newblock ``{DeepHunter: A Coverage-Guided Fuzz Testing Framework for Deep
  Neural Networks},''
\newblock in {\em ACM SIGSOFT International Symposium on Software Testing and
  Analysis (ISSTA)}, 2019.

\bibitem{issre18_mutation}
Lei Ma, Fuyuan Zhang, Jiyuan Sun, Minhui Xue, Bo~Li, Felix Juefei-Xu, Chao Xie,
  Li~Li, Yang Liu, Jianjun Zhao, and Yadong Wang,
\newblock ``{DeepMutation: Mutation Testing of Deep Learning Systems},''
\newblock in {\em The 29th IEEE International Symposium on Software Reliability
  Engineering (ISSRE)}, 2018.

\bibitem{du2019deepstellar}
Xiaoning Du, Xiaofei Xie, Yi~Li, Lei Ma, Yang Liu, and Jianjun Zhao,
\newblock ``Deepstellar: Model-based quantitative analysis of stateful deep
  learning systems,''
\newblock in {\em Proceedings of the 2019 27th ACM Joint Meeting on European
  Software Engineering Conference and Symposium on the Foundations of Software
  Engineering}, 2019, pp. 477--487.

\bibitem{ase18_gauge}
Lei Ma, Felix Juefei-Xu, Jiyuan Sun, Chunyang Chen, Ting Su, Fuyuan Zhang,
  Minhui Xue, Bo~Li, Li~Li, Yang Liu, Jianjun Zhao, and Yadong Wang,
\newblock ``{DeepGauge: Multi-Granularity Testing Criteria for Deep Learning
  Systems},''
\newblock in {\em The 33rd IEEE/ACM International Conference on Automated
  Software Engineering (ASE)}, 2018.

\end{thebibliography}

\end{document}



\linenumbers  %
\onecolumn
\linewidth\hsize \vskip 0.625in minus 0.125in \centering{
\title{BIAS FIELD POSES A THREAT TO DNN-BASED X-RAY RECOGNITION\\
Supplementary Material}
%
\name{%
\begin{tabular}{@{}c@{}}
Binyu Tian$^{1}$ \quad 
Qing Guo$^{2^{\ast}}$ \quad 
Felix Juefei-Xu$^{3}$ \quad
Wen Le Chan$^{2}$ \quad 
Yupeng Cheng$^{2}$ \\
Xiaohong Li$^{1^{\ast}}$ \quad
Xiaofei Xie$^{2}$\quad
Shengchao Qin$^{4}$\thanks{$^{\ast}$Qing Guo and Xiaohong Li are the corresponding authors (tsingqguo@ieee.org and xiaohongli@tju.edu.cn).}
\end{tabular}}
%

\address{$^{1}$College of Intelligence and Computing, Tianjin University, China \\ \quad $^{2}$Nanyang Technological University, Singapore \\
$^{3}$Alibaba Group, USA \quad $^{4}$Teesside University, UK}

\maketitle 
}


\justifying
In this material, we show and compare the visualization results of interpretable maps of three subject models, \ie, ResNet50, MobileNet, and DenseNet121, in Fig.~\ref{fig:analysis_resnet50}, \ref{fig:analysis_mobilenet}, and \ref{fig:analysis_densenet}, respectively. We first review the process of generating the interpretable map of each X-ray image with a given DNN model. Specifically, given an adversarial bias field example $\mathbf{X}^{\text{a}}$ generated by our method and the original image $\mathbf{X}$, we can calculate an interpretable map $\mathbf{M}$ for a deep neural network $\text{DNN}(\cdot)$ by optimizing
%
\begin{align}\label{eq:interp_map}
\argmin_{\mathbf{M}}~~\text{DNN}_y(\mathbf{M} \odot \mathbf{X}^\text{a} + (1-\mathbf{M}) \odot \mathbf{X}) \tag{8} \\~~
+ \lambda_{1} \| \mathbf{M}\|_{1} + \lambda_{2}\mathrm{TV} (\mathbf{M}) \nonumber
\end{align}
%
where $\text{DNN}_y(\cdot)$ denotes the score at label $y$ that is the ground truth label of $\mathbf{X}$ and $\mathrm{TV}(\cdot)$ is the total-variation norm. $\text{DNN}(\cdot)$ denotes the subject model we used and could be ResNet50, MobileNet, and DenseNet121, respectively. Intuitively, optimizing Eq.~(\ref{eq:interp_map}) is to find the region that causes misclassification of the DNN model. We optimize Eq.~(\ref{eq:interp_map}) via gradient decent in 150 iterations and fix $\lambda_{1}=0.05$ and $\lambda_{2}=0.2$. 

For each DNN, we can calculate an interpretable map for each X-ray image and get an averaging interpretable map. We show four examples and the averaging map for the three DNNs in Fig.~\ref{fig:analysis_resnet50}, \ref{fig:analysis_mobilenet}, and \ref{fig:analysis_densenet}, respectively and observe that: \ding{182} the interpretable maps calculated from different DNNs are almost the same, hinting that the bias-field-sensitive regions are independent to the DNN used for classification. \ding{183} the bias-field-sensitive regions principally locate at the top and bottom positions in all attacked images from ResNet50, MobileNet and DenseNet121. It all shows that the design of DNN needs to consider the spatial variations of xray pictures.

\renewcommand{\thefigure}{\Roman{figure}}

\begin{figure*}[!h]
	\centering
	\includegraphics[width=0.593\linewidth]{image/fig_anaylsis.pdf}
	\caption{Examples of exploring bias-field-sensitive regions. A subject model, \ie, {\bf ResNet50}, is employed to generate adversarial bias field examples for 240 X-ray images and we then use Eq.~\ref{eq:interp_map} to produce the interpretable map $\mathbf{M}$ for each image (\ie, the images at the second row where the maps are blended with the raw X-ray images for better understanding.). Finally, we can calculate an averaging map covering all interpretable maps and blend it with raw images (\ie, the images at the third row).}
	\label{fig:analysis_resnet50}
\end{figure*}

\begin{figure*}[!h]
	\centering
	\includegraphics[width=0.6\linewidth]{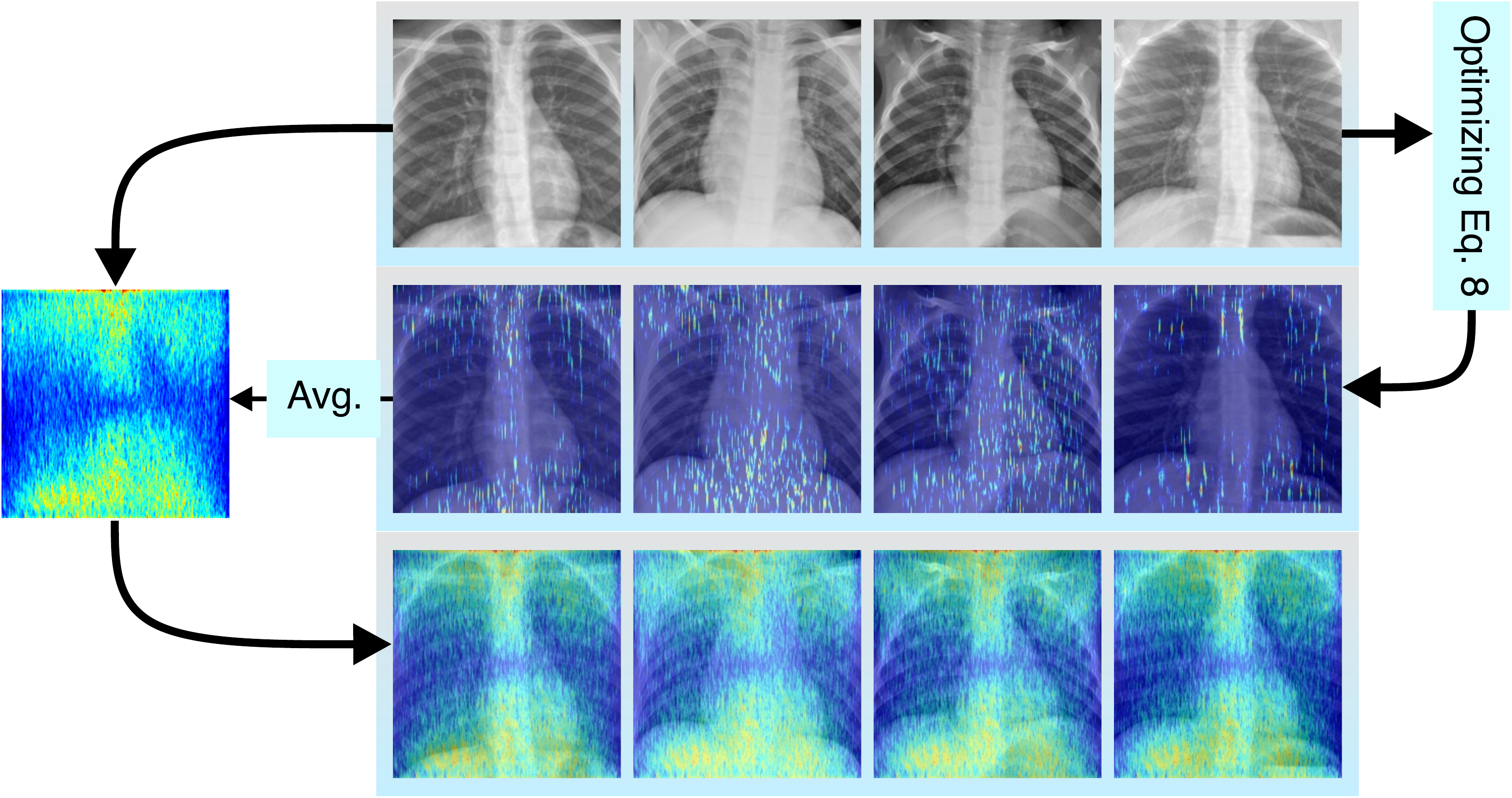}
	\caption{Examples of exploring bias-field-sensitive regions. A subject model, \ie, {\bf MobileNet}, is employed to generate adversarial bias field examples for 240 X-ray images and we then use Eq.~\ref{eq:interp_map} to produce the interpretable map $\mathbf{M}$ for each image (\ie, the images at the second row where the maps are blended with the raw X-ray images for better understanding.). Finally, we can calculate an averaging map covering all interpretable maps and blend it with raw images (\ie, the images at the third row).}
	\label{fig:analysis_mobilenet}
\end{figure*}
\begin{figure*}[!h]
	\centering
	\includegraphics[width=0.6\linewidth]{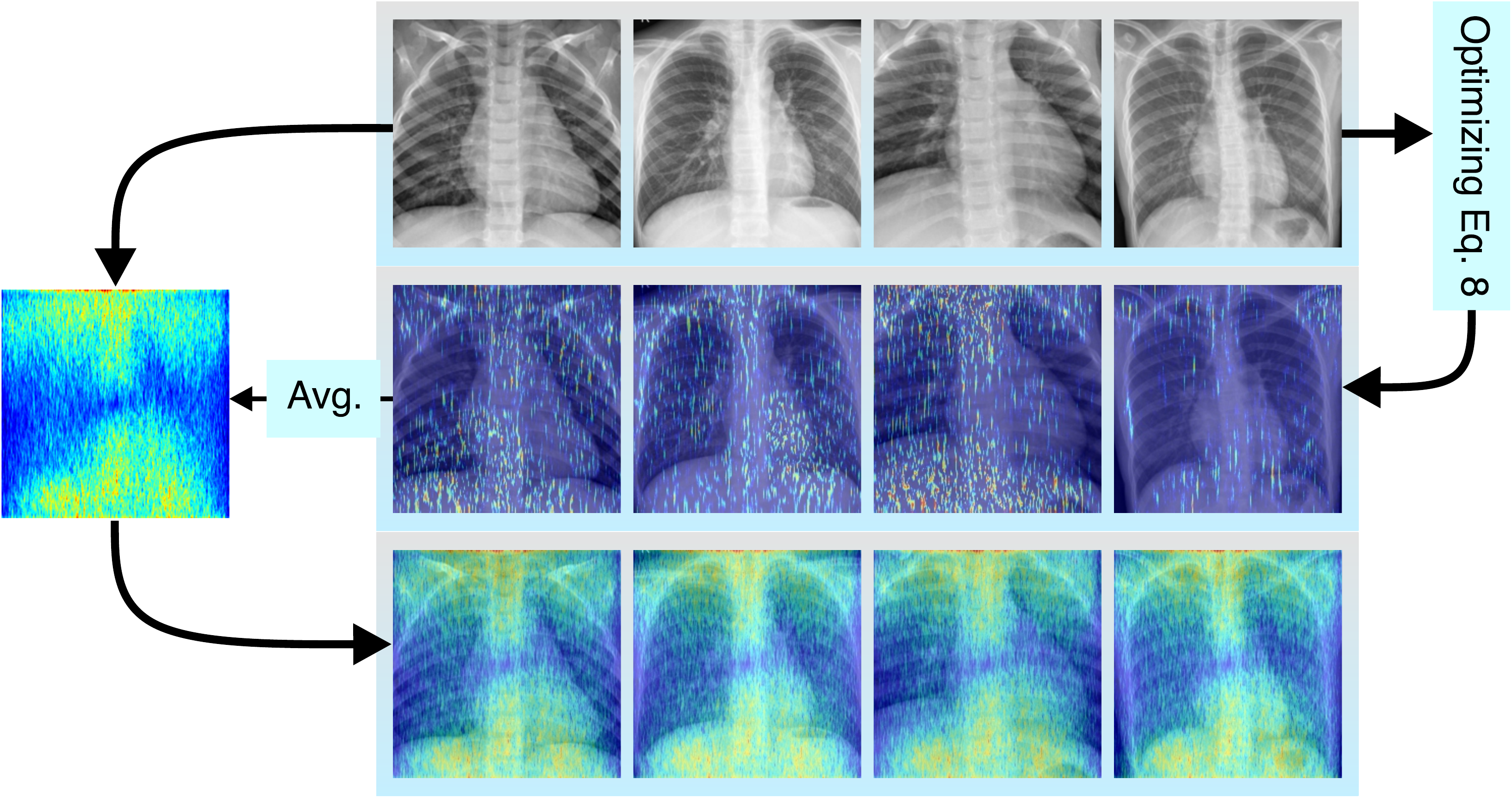}
	\caption{Examples of exploring bias-field-sensitive regions. A subject model, \ie, {\bf DenseNet121}, is employed to generate adversarial bias field examples for 240 X-ray images and we then use Eq.~\ref{eq:interp_map} to produce the interpretable map $\mathbf{M}$ for each image (\ie, the images at the second row where the maps are blended with the raw X-ray images for better understanding.). Finally, we can calculate an averaging map covering all interpretable maps and blend it with raw images (\ie, the images at the third row).}
	\label{fig:analysis_densenet}
\end{figure*}

\newpage